\documentclass[jcp,amsmath,amssymb,amsfonts,superscriptaddress,preprint]{revtex4-1}

\usepackage{graphicx}
\usepackage{natbib}

\usepackage{verbatim}

\begin{document}

\title{\textbf{Photon echo in exciton-plasmon nanomaterials: a time-dependent signature of strong coupling}}

\author{Adam Blake}
\affiliation{Department of Physics, Arizona State University, Tempe, Arizona 85287}
\author{Maxim Sukharev}
\email{maxim.sukharev@asu.edu}
\affiliation{Department of Physics, Arizona State University, Tempe, Arizona 85287}
\affiliation{College of Integrative Sciences and Arts, Arizona State University, Mesa, Arizona 85212}

\date{\today}
\begin{abstract}
We investigate the dynamics of photon echo exhibited by exciton-plasmon systems under strong coupling conditions. Using a self-consistent model based on coupled Maxwell-Bloch equations we investigate femtosecond time dynamics of ensembles of interacting molecules optically coupled to surface plasmon supporting materials. It is shown that observed photon echoes under two pulse pump-probe sequence are highly dependent on various material parameters such as molecular concentration and periodicity. Simulations of photon echoes in exciton-plasmon materials reveal a unique signature of the strong exciton-plasmon coupling, namely a double-peak structure in spectra of recorded echo signals. This phenomenon is shown to be related to hybrid states (upper and lower polaritons) in exciton-plasmon systems under strong coupling conditions. It is also demonstrated that the double-peak echo is highly sensitive to mild deviations of the coupling from resonant conditions making it a great tool for ultrafast probes. 
\end{abstract}
\maketitle

\section{Introduction}
\label{sec:Intro}

The research field of nanoplasmonics has grown substantially in past few years due to tremendous progress in fabrication techniques and optical characterization \cite{Zayats2005131,Stockman:11,doi:10.1021/cr200061k}. Due to strong spatial localization of electromagnetic modes corresponding to surface plasmon-polariton resonance in such materials, one may investigate the fundamentals of light-matter interaction on a single molecule level \cite{Chikkaraddy:2016aa}. A new type of nanomaterials with molecular excitons strongly optically coupled to plasmons is also extensively discussed\cite{Torma}. While the linear plasmonics is still enjoying significant interest \cite{ADMA:ADMA200700678}, the field of nonlinear plasmonics \cite{Kauranen:2012aa} recently began to attract considerable attention \cite{Butet:2015aa}. An obvious extension of various nonlinear spectroscopy techniques\cite{MukamelBook} to the domain of plasmonics is also an ongoing effort \cite{Vasa2013,Sukharev2013,Metzger:2016aa}.

Free induction decay is a well-documented phenomenon that was first demonstrated in nuclear magnetic resonance and is also observed in optics \cite{Eberly}. A population of quantum emitters can exist in which all emitters have the same \textit{central} transition energy but each is detuned by some amount from the central transition energy due to inhomogeneous broadening \cite{PhysRev.141.391}, which can result from conditions such as Doppler shift in individual gas molecules or variations in electric field from point to point in solids \cite{Mandel}. The macroscopic polarization of an ensemble of emitters is the sum of each individual contribution from every emitter in the ensemble. Given that the transition energies of the emitters in the ensemble are described by a distribution, each individual emitter oscillates at a frequency that is slightly different from the others after the system is pumped by a strong incident pulse. As a result, all of the emitters begin to oscillate in phase at first, but eventually they dephase within a characteristic inhomogeneous lifetime and, if left alone, never re-phase again \cite{Eberly}. However, for times less than the natural lifetime of the emitter, each emitter is still oscillating. One can in principle invert the dephasing process by applying a second pump in the form of a $\pi$ pulse. The oscillations then all \textit{run in reverse}, resulting in a subsequent rephasing. The ensemble exhibits a non-zero macroscopic polarization once again eventually emitting radiation known as a photon echo signal.

This technique is widely used in chemistry and is referred to as photon echo spectroscopy \cite{Boeij1998}. Inhomogeneous effects due to variations in an emitter's surroundings cause each emitter to oscillate at a slightly different frequency than the others, and photon echo spectroscopy \textit{removes} this effect. Any remaining dephasing is irreversible by the echo technique, and is revealed as diminished intensity of the echo \cite{Cho1992}. For example, as the delay in applying the $\pi$ pulse increases, the natural lifetime of the emitters causes all of their oscillations to decrease, resulting in an echo with lower intensity. Thus a time and frequency structure of a detected photon echo contains important information about a probed system.

Additionally, the recovery of a time signal after dephasing offers prospects for memory storage. In \cite{Langer2014}, the optical properties are copied to a spin system whose lifetime is much longer than that of the optical system thus extending the duration of the system's memory.  

Our major interest in this paper is to test a concept of photon echo in strongly coupled exciton-plasmon systems in order to understand any new features unique to the strong coupling regime. First, we briefly overview the numerical model used. Next, we consider the application of a $\pi/2$ pulse to a 1-D ensemble followed by application of a $\pi$ pulse generating a photon echo signal. The density of molecules in this ensemble is varied and the strength of the echo is considered in terms of the transmission and reflection of the ensemble. Finally, we investigate the photon echo of two exciton-plasmon systems, namely a periodic array of slits and a core-shell nanoparticle, each combined with inhomogeneously broadened molecules.

\section{Model}
\label{sec:Model}
The time dynamics of molecules is described by the Liouville equation
\begin{equation}
i\hbar\frac{d\hat{\rho_k}}{dt}=[\hat{H}_k,\hat{\rho_k}]-i\hbar\hat{\Gamma}\hat{\rho_k},
 \label{Bloch}
\end{equation}
where $k$ corresponds to the index of a specific transition energy, $\hbar \omega_{0k}$, of a given molecule at a given spatial position, $\hat{\rho_k}$ is the single-molecule density matrix, $\hat{\Gamma}$ describes relaxation processes, and $\hat{H_k}$ is the Hamiltonian of a single molecule with a dipole moment operator $\hat{\vec{\mu}}_k$ interacting with a local electric field
\begin{equation}
\hat{H}_k=\hat{H}_{0k}-\hat{\vec{\mu}}_k\cdot\vec{E},
\end{equation}
here the field-free Hamiltonian for a molecule with energy index $k$ is $\hat{H_0}=\hbar \omega_{0k} \left | k \right > \left < k \right |$.
The dynamics of the electric field, $\vec{E}$, is governed by corresponding Maxwell's equations as discussed below.

We consider an ensemble of two-level molecules explicitly including the inhomogeneous broadening (IB) described by the Gaussian distribution $G_k$
\begin{equation}
 G_k=G_0\exp\left({\frac{-(\omega_{0k}-\omega_C)^{2}} {2\Delta\omega^{2}}}\right),
 \label{Gaussian}
\end{equation}
where $G_0$ is the normalization constant, $\omega_C$ is the energy corresponding to the maximum in the Gaussian distribution, and $\Delta\omega$ characterizes the full width at half-maximum (FWHM) of the distribution, which describes how broad a given IB is. Specific values of $\Delta \omega$ in this work are: "narrow" is $0.136$ eV, "intermediate" is $0.188$ eV, and "broad" is $0.236$ eV. The distribution (\ref{Gaussian}) is normalized such that
\begin{equation}
 \sum_{k} G_k = 1.
 \label{GaussNorm}
\end{equation}

The expectation value of the dipole moment is calculated as a sum over all molecular transition frequencies
\begin{equation}
 \left<\vec{\mu}\right>=\sum_{k}G_k\text{Tr}(\hat{\rho}\hat{\vec{\mu}}_k)
 \label{dipole}
\end{equation}
The macroscopic polarization, $\vec{P}$, is calculated according to $\vec{P}=n_a\left<\vec{\mu}\right>$, where $n_a$ is the number density of molecules. The polarization current, $\vec{J_p}=d\vec{P}/  dt$, is subsequently inserted into Maxwell's equations, the solutions of which at a give time and spatial positions are then used to update macroscopic polarization. The following set of parameters describing a molecule is used in this paper: the transition dipole moment is $10$ Debye, the radiationless lifetime of the excited state is $1$ ps, and the pure dephasing time is $800$ fs. 

To model electromagnetic (EM) wave propagation and its interaction with molecules and metal in time and space in addition to the Bloch equations (\ref{Bloch}) we numerically integrate Maxwell's equations employing the finite-difference time-domain method (FDTD) \cite{Taflove}
\begin{subequations}
 \begin{eqnarray}
\frac{\partial \vec{B}}{\partial t}&=&-\nabla\times\vec{E}, \label{Faraday}\\
\epsilon_0\epsilon_r\frac{\partial \vec{E}}{\partial t}&=&\frac{1}{\mu_0}\nabla\times\vec{B} - \vec{J}, \label{AmpereMaxwell}
 \end{eqnarray}
\end{subequations}
where $\mu_0$ and $\epsilon_0$ are the permeability and the permittivity of free space, respectively, $\vec{B}$ is the magnetic field, $\vec{J}$ corresponds to the current density in spatial regions occupied by metal and polarization current in regions filled with molecules, $\epsilon_r$ is the relative permittivity at a given spatial position (1 for vacuum, see below). 

To account for the material dispersion of metal we employ the Drude model
\begin{equation}
\epsilon(\omega)=\epsilon_r - \frac{\omega_p^2}{\omega^2-i\gamma\omega}, \label{DrudePermittivity}
\end{equation}
where $\omega_p$ is the plasma frequency, $\gamma$ is the phenomenological damping, and $\epsilon_r$ is the high-frequency limit of the dielectric function. For silver, we use the following parameters \cite{Gray2003}: $\hbar\omega_p = 11.59$ eV, $\hbar\gamma = 0.2027$ eV, and $\epsilon_r = 8.26$. The corresponding time dynamics of the current density $\vec{J}$ satisfies the following equation \cite{Taflove}
\begin{equation}
\frac{\partial \vec{J}}{\partial t}=-\gamma\vec{J}+\epsilon_0\omega_p^2\vec{E}.
 \label{J}
\end{equation}
Coupled equations (\ref{Bloch}), (\ref{Faraday}), (\ref{AmpereMaxwell}), and (\ref{J}) are solved self-consistently using home-built codes. The numerical convergence is achieved for the grid with $\delta x = 1$ nm and a time step of $\delta t = 1.67$. 

It should be noted that inevitable discretization of the transition energy distribution (\ref{Gaussian}) results in unphysical rephasing which must be carefully monitored in all simulations in order to avoid spurious numerical artifacts. In brief, after initial excitation of the molecular system, free induction decay occurs followed by a numerical rephasing after a delay which is proportional to the $1/\delta\omega$, where $\delta\omega$ is the discretization step in (\ref{Gaussian}). The rephasing is obviously unphysical for a realistic number of molecules (i.e. continuous transition frequency distribution with the revival time being infinite \cite{Eberly}). In all our simulations we use an energy step $\hbar\delta\omega$ no larger than $0.01$ eV, corresponding to a revival time of $400$ fs or longer.

\begin{figure}
\centering
\includegraphics[width=\textwidth]{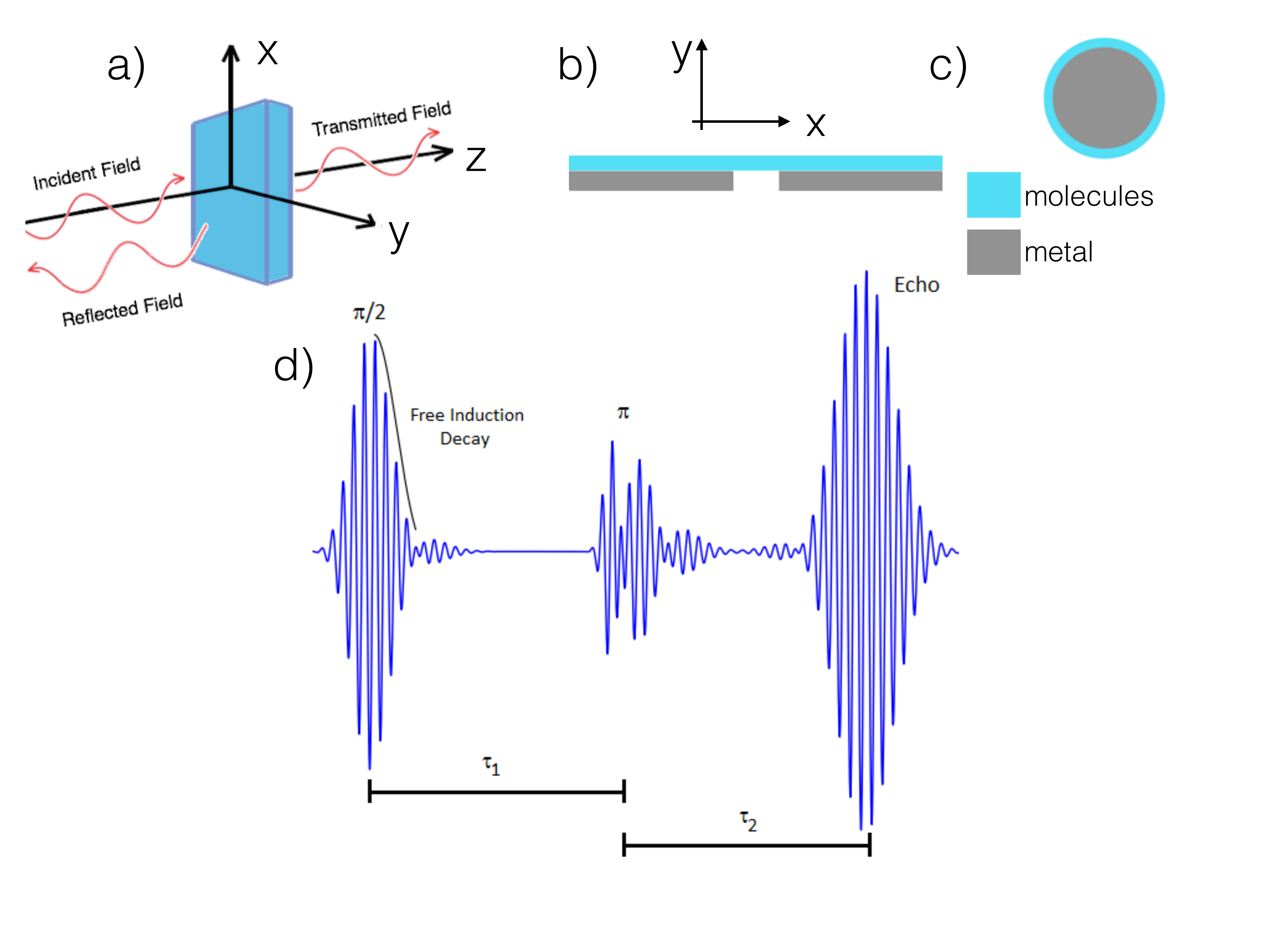}
\caption{\label{figure1} Panels (a) - (c) are schematic setups of systems considered: (a) an extended 1-D ensemble of interacting molecules with a finite thickness along $z$ and infinite in $x$ and $y$ (the incident field propagates in $z$ direction and is polarized along $x$); (b) the periodic array of slits coupled to a thin film of molecules placed on the input side (here the incident field is propagated from top to bottom and is polarized in horizontal direction); (c) the core-shell nanoparticle consisting of a metallic core surrounded by a thin shell of interacting molecules. Panel (d) shows the schematics of the two pulse $\pi/2$-$\pi$ pump-probe sequence with the subsequent photon echo.}
\end{figure}

\section{Results and Discussion}
\label{sec:Results}
First we consider noninteracting molecules driven by linearly polarized incident field, the polarization current $dP/dt$ calculated in the direction of the incident field polarization is used as an indication of rephasing. We employ the two-pulse pump-probe photon echo sequence as illustrated in Fig. \ref{figure1}d. $\tau_1$ corresponds to the time between the maximum of the $\pi/2$ pulse and the $\pi$ pulse, and $\tau_2$ corresponds to the time between the maximum of the $\pi$ pulse and the maximum of the echo. These times are roughly equal as they should be \cite{Mandel}. The duration and amplitude of the $\pi/2$ pulse are such that it drives the density matrix element $\rho_{11}$ of a molecule (whose transition energy corresponds to $\omega_C$) to $0.5$. The $\pi$ pulse completely inverts the molecule. Molecules are excited by a $\pi/2$ pulse polarizing the sample, which then undergoes dephasing resulting in observed free induction decay. A $\pi$ pulse is then applied, and the maximum of the photon echo signal is observed at time $\tau_1 + \tau_2$. Using these simulations as a test case we verified our numerical procedure and proceed to the case of interacting molecules.

The ensemble considered thus far has been a collection of noninteracting molecules. In order to determine the influence of collective effects on the echo amplitude and its form, we employ Maxwell-Bloch formalism applied to the one-dimensional thin molecular layer as depicted in Fig. \ref{figure1}a. By integrating coupled Maxwell-Bloch equations we allow molecules to interact via emission and absorption of EM radiation. We first examined the main results of Ref. [\onlinecite{PhysRevA.91.043835}] for the molecules with IB calculating the transmission, $T$, and reflection, $R$ (not shown). Both $T$ and $R$ expectedly broaden and flatten as the inhomogeneous broadening is introduced and further increased. The transmission decreases toward $0$ and the reflection increases toward $1$ as the density increases, both are expected results as at high densities the electric susceptibility near central frequency $\omega_C$ becomes negative indicating the fact that the molecules oscillate out-of-phase with respect to the incident field thus efficiently canceling it out inside the sample \cite{Raiju2014}. Distortion is observed on the edges of the spectral features due to increasing interference between the reflected signals from varying depths of the ensemble.

\begin{figure}
\begin{center}
\includegraphics[width=\textwidth]{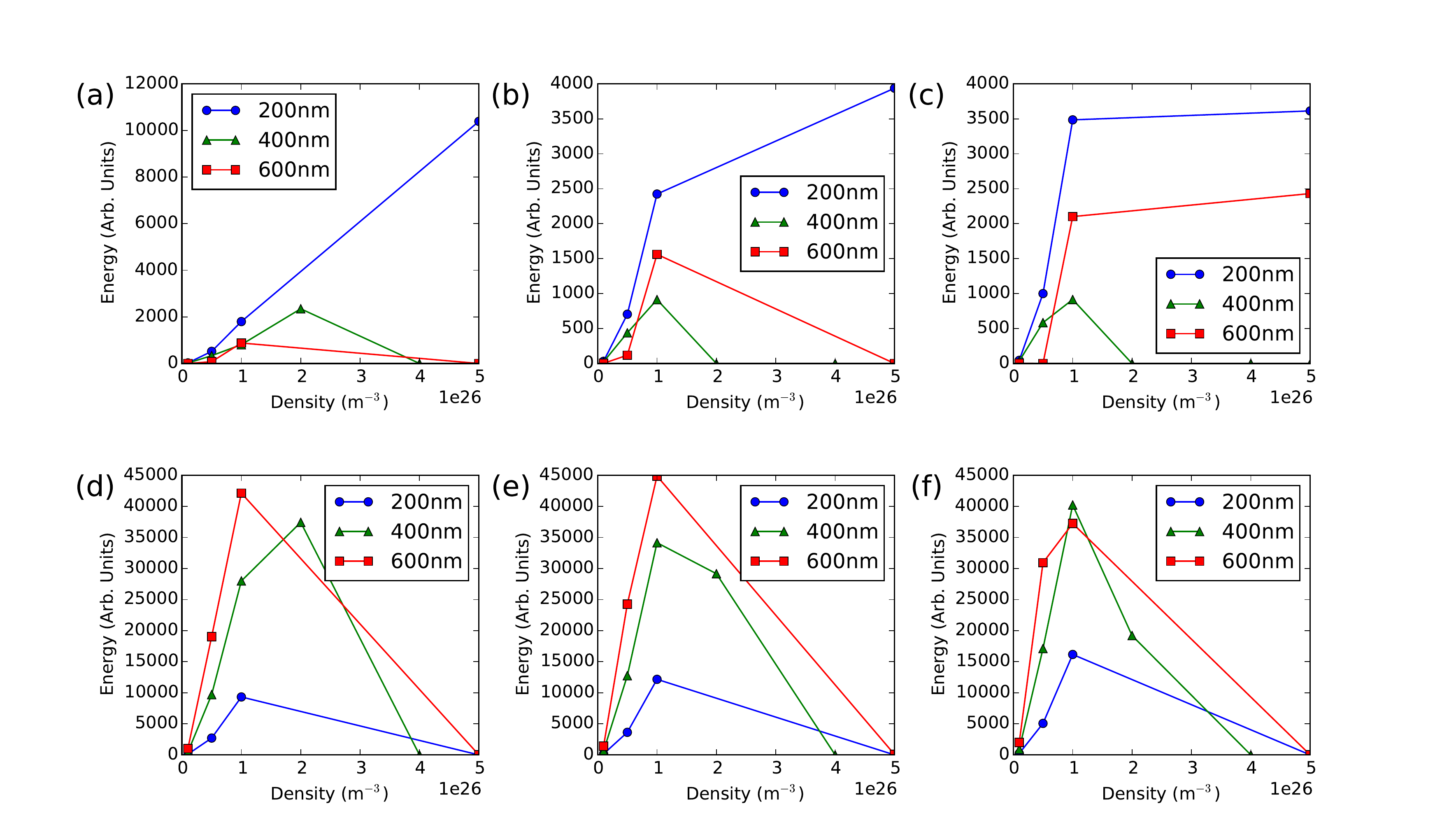}
\caption{\label{figure2} EM energy of the photon echo signal on the input and output sides of a 1-D ensemble of interacting molecules as a function of the molecular density. Results for the molecular layer of the thickness of $200$ nm are shown as circles, triangles correspond to the thickness of $400$ nm, and squares correspond to the thickness of $600$ nm. The signal increases with density at lower densities and in several cases drops sharply at higher densities, which is attributed to highly increased reflection. Different panels correspond to different detection sides (output or input) and different FWHM of the inhomogeneous broadening distribution (\ref{Gaussian}). Energy detected on the input side is shown in panels (a) ($\Delta \omega=0.236$ eV), (b) ($\Delta \omega=0.188$ eV), and (c) ($\Delta \omega=0.136$ eV). The detection on the output side is shown in panels (d) ($\Delta \omega=0.236$ eV), (e) ($\Delta \omega=0.188$ eV), and (f) ($\Delta \omega=0.136$ eV).}
\end{center}
\end{figure}

We apply the $\pi/2$-$\pi$-pulse sequence tuned to the central frequency of a single molecule set at $\omega_C=2.0$ eV, then record the energy of the echo signal. The signal is calculated as an integral of the electric field intensity over time on both the input (reflection) and output (transmission) sides of the molecular layer. Simulations are performed for several densities, molecular layer thicknesses, and values of inhomogeneous broadening. The results are shown in Fig. \ref{figure2}. On both sides of the ensemble, the echo amplitude increases with the density for lower densities but, in many instances, decreases sharply at higher densities. This can be understood in terms of how the transmission and reflection of the ensemble vary with density. At higher densities, the reflection approaches $1$ and the transmission becomes very small as discussed above. Thus, the driving fields do not make it as far into the ensemble for higher densities making photon echo signal lower as not all molecules completely re-phase and therefore reducing the macroscopic polarization, which in turn reduces the intensity of photon echo. Another interesting observation is that there is obviously an optimal molecular density, at which the energy of photon echo detected on the output side has a global maximum. This is explained below.

\begin{figure}
\begin{center}
\includegraphics[width=\textwidth]{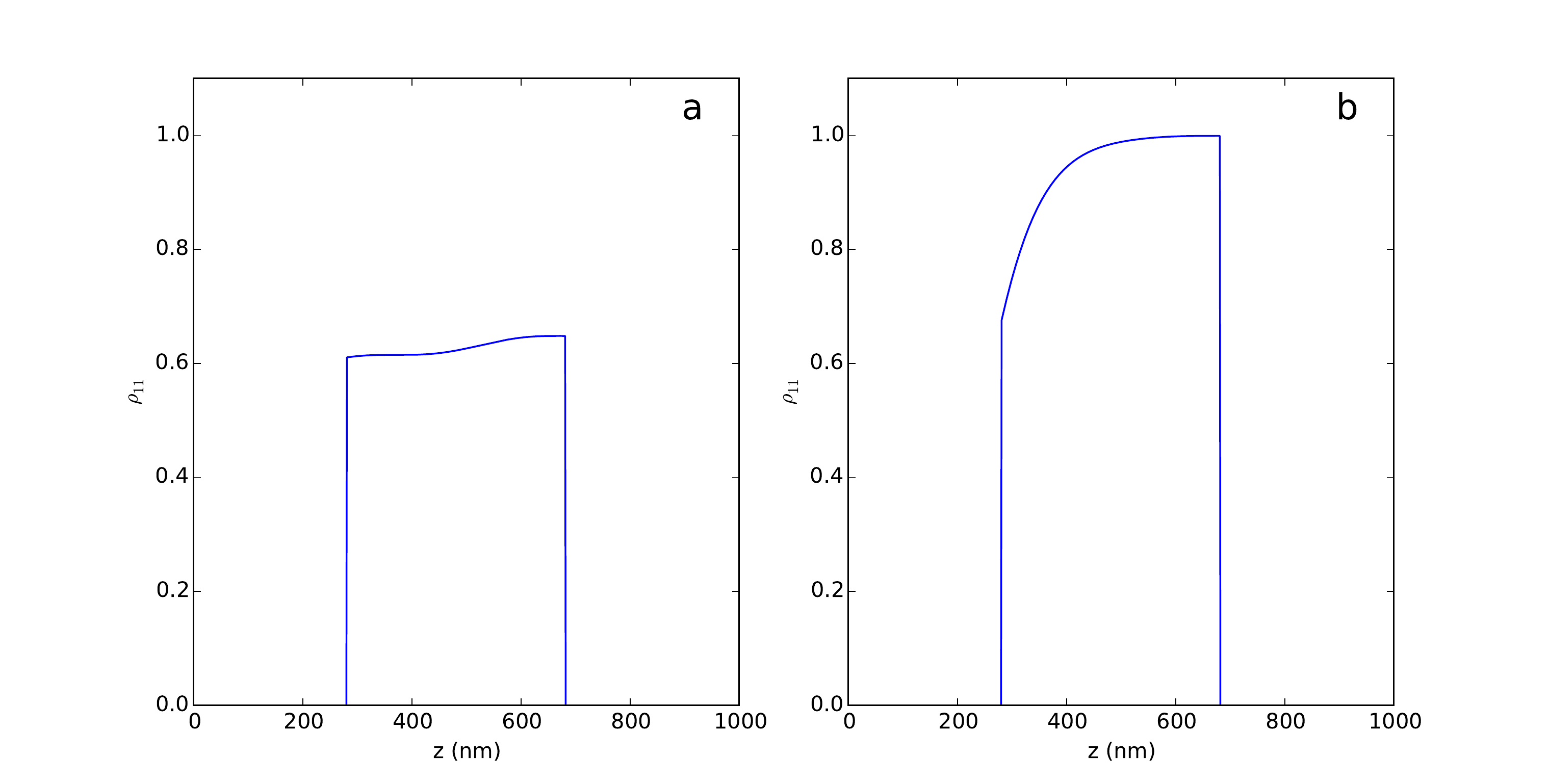}
\caption{\label{figure3} Ground state population as a function of the longitudinal coordinate along the 1-D ensemble of molecules calculated for $\Delta \omega=0.236$ eV at the end of the $\pi/2$ pulse. The molecular layer has a thickness of $400$ nm. Two panels correspond to different molecular concentrations: (a) $10^{25}$ m$^{-3}$, (b) $10^{26}$ m$^{-3}$}
\end{center}
\end{figure}

A snapshot of the molecular population as a function of the position is shown in Fig. \ref{figure3}. The pulse travels from left to right, and the ground state population along the ensemble is calculated immediately after the $\pi/2$ pulse passes through the ensemble. It is seen that, for a higher density, the ground state population ${\rho}_{11}$ approaches $1$ as we look farther into the ensemble; the local EM fields are smaller and therefore do not drive the molecules through complete $\pi/2$ cycle. With this in mind, it makes sense that the echo on the output side drops off at higher densities: more and more of the ensemble (toward the output side) ceases to participate in the echo at higher densities, and the echo from the input side is not able to travel through the ensemble to the output side. Thus, the echo increases with increasing density until reflection increases significantly, disrupting the echo process throughout the ensemble. The input side is somewhat more complicated, as there is obvious interference between incoming and outgoing reflected waves, and the reflected waves from lesser depths interfere with reflected waves from deeper regions of the ensemble. Furthermore, the ensemble is driven by high fields on resonance, leading to higher dispersion and as a result the low group velocity. The latter results in high spatially modulated populations which obviously have a significant effect on the generation of a photon echo \cite{Blake2015}.

\begin{figure}
\begin{center}
\includegraphics[width=\textwidth]{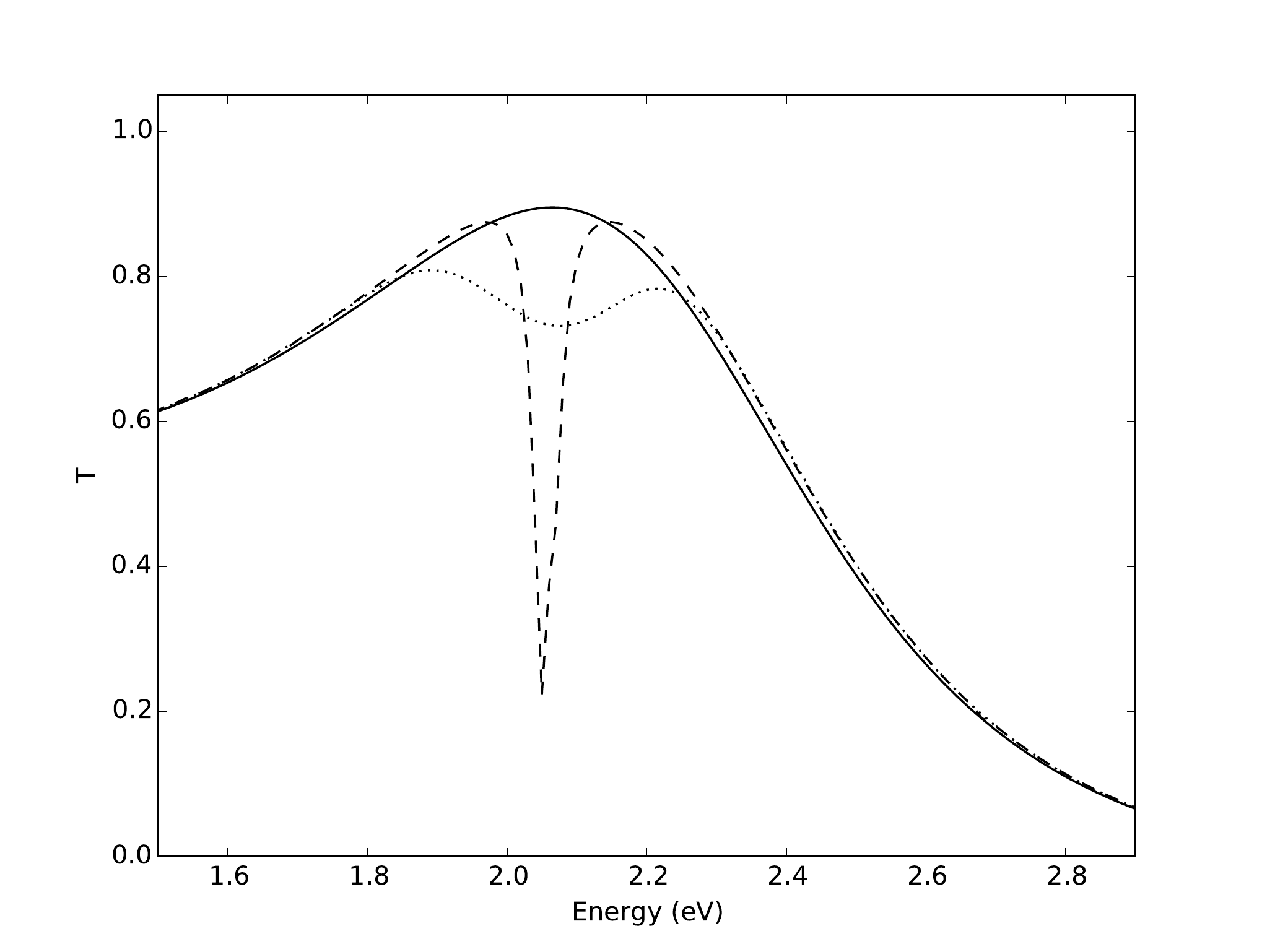}
\caption{\label{figure4} Linear transmission through the periodic array of slits as a function of the incident photon energy. The solid line indicates transmission calculated for the bare metal slits, the dashed line is for a hybrid system with no inhomogeneous broadening, and the dotted line corresponds to a hybrid system with inhomogeneous broadening corresponding to $\Delta \omega = 0.236$ eV. The period is $320$ nm, metal film thickness is $200$ nm, the slit width is $140$ nm, the molecular layer has a thickness of $20$ nm, and the molecular number density is $10^{26}$ m$^{-3}$.}
\end{center}
\end{figure}

We now turn our attention to an exciton-plasmon system comprising the periodic array of slits and molecules with inhomogeneous broadening as schematically depicted in Fig. \ref{figure1}b. The bare metallic grating exhibits a set of surface plasmon-polariton (SPP) resonances in the form of enhanced transmission and decreased reflection \cite{0034-4885-72-6-064401}. For the parameters used we consider the SPP resonance at $2.06$ eV as our working mode. The corresponding spatial distribution of EM field is highly localized near slit edges on both input and output sides of the array \cite{Weiner:11}. The resonant molecules (whose transition energy is $2.06$ eV) with no inhomogeneous broadening placed inside a thin layer on top of the array lead to the Rabi splitting of the SPP mode into an upper and lower polariton indicating the strong coupling regime between the SPP field and molecules. However when the molecules exhibit inhomogeneous broadening, the Rabi splitting is still observable but it is significantly broader as one may expect as not all molecules are as strongly coupled compared to the previous case. Nonetheless anticipated broader spectrum one can actually observe significantly higher Rabi splitting. Fig. \ref{figure4} shows linear transmission spectra for bare slits, the hybrid system with no inhomogeneous broadening, and for the hybrid system whose molecules are inhomogeneously broadened. As seen from Fig. \ref{figure4}, the Rabi splitting for the hybrid system with no inhomogeneous broadening reaches $180$ meV whereas that for a hybrid system with inhomogeneous broadening is $325$ meV. This is a surprising result. One would have expected to observe the opposite behavior of the Rabi splitting. However the increase of the splitting in systems with distributed energies is not uncommon. Simulations of optics of exciton-plasmon systems comprised diatomic molecules (described using the full machinery of a ro-vibrational structure for each electronic state) and metal reveal similar tendency \cite{EricMaxim_molecules}. The increase of the exciton-plasmon coupling is observed for the molecules described by two potential energy surfaces with the excited one being a dissociative state, i.e. having a continuum of ro-vibrational states. This in part can be explained by the fact that plasmon resonances are usually quite broad and a wide range of molecular transition energies $\omega_{0k}$ contribute coherently to the coupling. It should also be noted that in our simulations the pure dephasing time is the same for all $\omega_{0k}$ resulting in the same characteristic decay time for all molecules.

\begin{figure}
\begin{center}
\includegraphics[width=\textwidth]{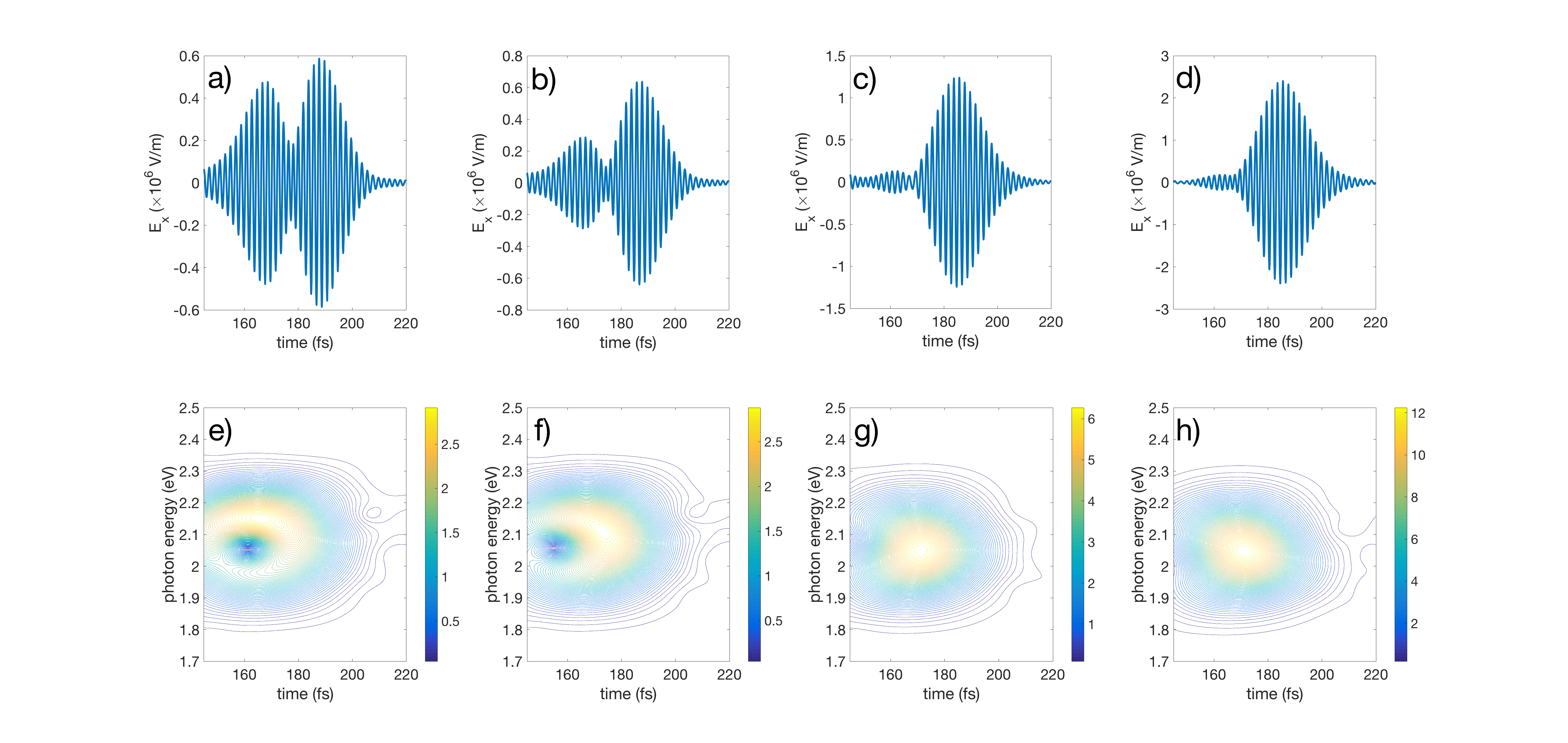}
\caption{\label{figure5} The upper row (panels (a) - (d)) shows the time dependent photon echo signals observed in the exciton-plasmon nanosystem comprising periodic array of slits and molecules. The lower row (panels (e) - (h)) shows the corresponding Husimi transformations of the photon echo plotted as functions of time and frequency. A nondispersive spacer of varying thickness is inserted between the film and the slits, and the double-peaked structure gradually reduces to a single peak as the spacer thickness is increased. The spacer length is (a, e) $0$ nm, (b, f) $5$ nm, (c, g) $25$ nm, and (d, h) $75$ nm. The molecular density is $10^{26}$ m$^{-3}$, the thickness of the molecular film is $20$ nm, $\Delta \omega = 0.236$ eV, the period of the slit array is $320$ nm, and the metal film thickness is $200$ nm.}
\end{center}
\end{figure}

\begin{figure}
\begin{center}
\includegraphics[width=\textwidth]{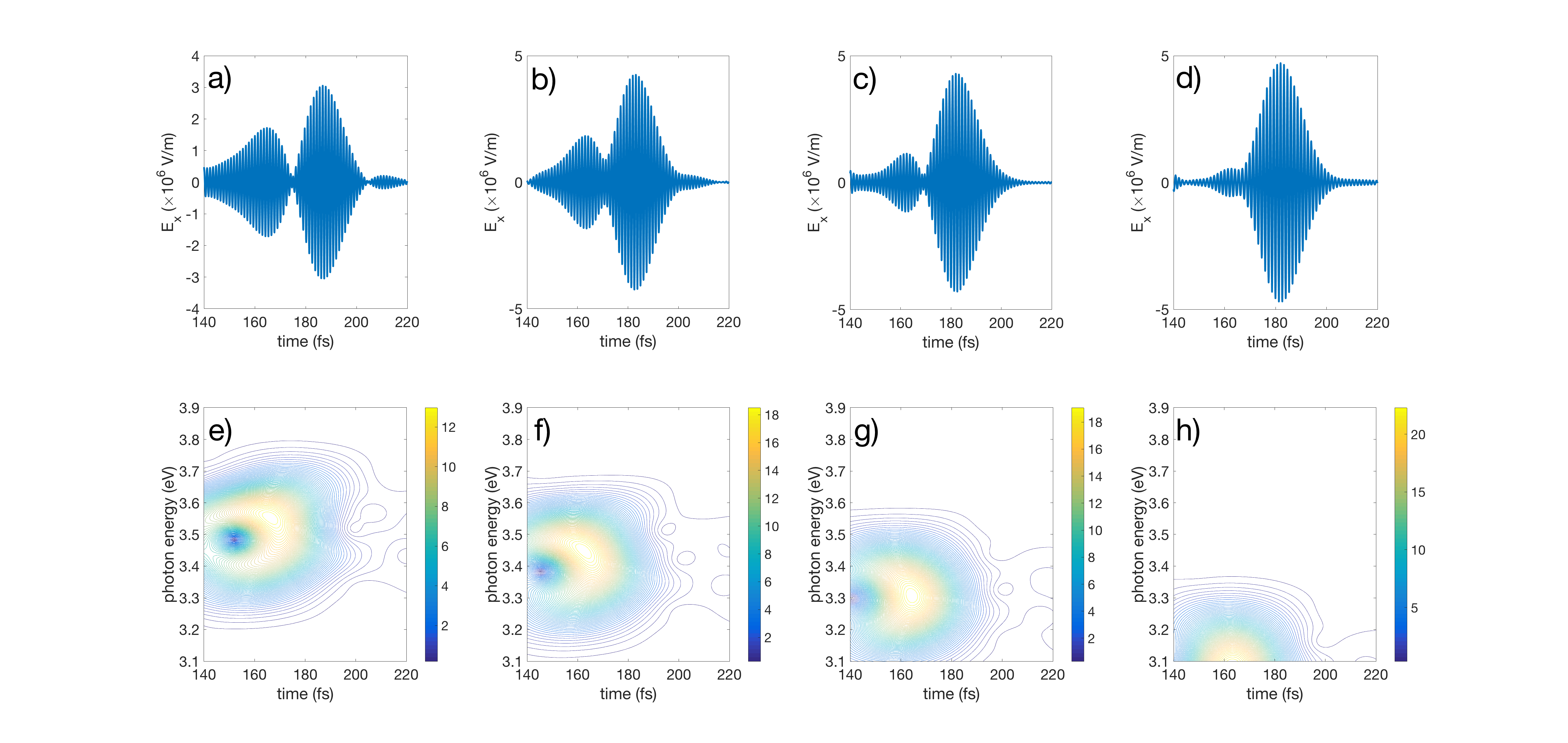}
\caption{\label{figure6} The upper row (panels (a) - (d)) shows the time dependent photon echo signals observed for the exciton-plasmon nanosystem comprising a metal core and a molecular shell. The lower row (panels (e) - (h)) shows the corresponding Husimi transformations of the photon echo plotted as functions of time and frequency. The central molecular transition energy, $\omega_C$, is (a, e) $3.50$ eV, (b, f) $3.40$ eV, (c, g) $3.30$ eV, and (d, h) $3.09$ eV.
The material of the core is silver with a radius of $25$ nm and the molecular shell with a thickness of $20$ nm consists of interacting molecules with inhomogeneous broadening ($\Delta \omega = 0.236$ eV). The localized surface plasmon resonance of the bare core is $3.59$ eV.}
\end{center}
\end{figure}

We now apply the $\pi/2$-$\pi$-pulse sequence to the array of slits with molecules on top and record the transverse component of the electric field on both the input and output sides of the system for the extended period of time when the echo is observed. We set central transition frequency $\omega_C$ to the SPP mode at $2.06$ eV. What is seen is very different from the conventional photon echo generated by a 1-D ensemble: the observed echo has a clear double-peaked structure. The frequency-time analysis of the echo signal reveals two peaks as seen in Fig. \ref{figure6}. We proceed to show that the double-peaked structure has its origins in the interaction between the SPP mode and molecules.

The SPPs are spatially localized to within tens of nanometers of a metal-dielectric interface. In order to ascertain whether our observations of the double peak echo is influenced by surface plasmons, we insert a spacer layer of variable thickness between the molecular film and the slits. As the spacer thickness is increased, the film is moved into progressively weaker plasmon fields leading to a noticeably lower molecule-plasmon coupling. As shown in Fig. \ref{figure5} (panels (a) - (d)), the double-peaked structure of the echo gradually disappears as the spacer thickness is increased. The amount of coupling between plasmons and molecules depends, among other things, on the strength of the plasmon fields. We therefore expect that if the double-peaked structure of the echo is caused by the strong coupling between plasmons and molecules, a spacer layer of increasing size will reduce the echo to a single peak, which is exactly what is observed. It is interesting to see how sensitive the double-peaked photon echo is to the molecule-plasmon coupling as it nearly vanishes for the spacer with a thickness above $30$ nm. Even though plasmon fields are still quite significant at distances of $25$ nm above the metal the spectrum of echo is already reduced to a single resonance at $\omega_C$. The corresponding Husimi transformations \cite{Brixner:01} of the time signals (Fig. \ref{figure5} (panels (e) - (h))) further reveal a clear contribution from the hybrid states formed due to strong coupling between molecules and the SPP mode (upper and lower polariton states) to the time dynamics of the photon echo. When the molecular layer is placed right on top of the metal array an obvious signature of two polaritonic branches appears in the time-frequency map with maxima near $1.95$ eV and $2.15$ eV (compare those with maxima seen in the linear transmission, dotted line in Fig. \ref{figure4}). As the distance between the molecular layer and metal increases the Rabi splitting gradually decreases eventually leading to a single peak in photon echo as the damping rates begin to surpass the coupling strength between molecules and the SPP mode. We thus conclude that the double peak structure observed in photon echo signals is a unique signature of the upper and lower polaritons that are formed as a result of the strong coupling between SPPs and molecules.

To further confirm that the observed double-peaked photon echo signals are exhibited by exciton-plasmon systems under strong coupling conditions we consider a core-shell nanoparticle with a metal core and a shell comprised molecules with inhomogeneous broadening as schematically depicted in Fig. \ref{figure1}c. The bare silver core has a localized plasmon-polariton resonance at $3.59$ eV as revealed in calculations of the scattering. We then add resonant molecules with the frequency distribution (\ref{Gaussian}) centered at $3.59$ eV with $\Delta \omega = 0.236$ eV. Simulations of the linear response of the core-shell particle show the expected Rabi splitting with hybrid modes at $3.4$ eV and $3.6$ eV. In order to tune the molecule-plasmon coupling we vary the central energy of the distribution of molecules, $\omega_C$. By detuning $\omega_C$ from the plasmon resonance of the metal core we alter the Rabi splitting. It is expected that the double-peaked photon echo will gradually change its form to a single peak. Fig. \ref{figure6} shows the results of simulations demonstrating the gradual loss of the double-peaked structure in the echo. The smaller Rabi splitting with larger detuning leads to a single dominant frequency in corresponding Husimi maps. We also performed simulations with variable detuning applied to the hybrid slits system, both by adjusting the molecules' central energy as well as adjusting the slit period (and therefore altering the surface plasmon energy). As expected, the double-peaked structure vanished in each instance as the detuning was increased.

\section{Conclusion}
\label{sec:Conclusion}
We performed rigorous numerical studies of photon echo in exciton-plasmon systems under strong coupling conditions using Maxwell-Bloch formalism. It is shown that photon echoes are highly dependent on materials parameters such as a thickness of molecular layer and molecular density. We have demonstrated that our numerical methods reproduce conventional photon echo signals. We have characterized the echo from a one-dimensional ensemble of interacting molecules in terms of density, ensemble thickness, and the amount of inhomogeneous broadening. It is demonstrated that in one-dimensional systems the echo detected on the output side has an optimal molecular concentration at which the transmitted signal has a maximum energy. To the best of our knowledge it is shown for the first time that applying two-pulse photon echo sequence to exciton-plasmon systems leads to double-peaked echoes. This is demonstrated by different means to be of plasmonic origin, with the double-peaked structure resulting from the hybrid modes of the system (upper and lower polaritons). It is also shown that the unique signature of upper and lower polaritons in photon echoes is highly sensitive to exciton-plasmon coupling making it a great tool for ultrafast optical probes of nanomaterials to further scrutinize fundamentals of light-matter interaction. 

\section{Acknowledgments} 
This work is supported by the Air Force Office of Scientific Research under grant No. FA9550-15-1-0189. The authors are also grateful to the financial support provided by Binational Science Foundation under grant No. 2014113.


\begin{thebibliography}{27}%
\makeatletter
\providecommand \@ifxundefined [1]{%
 \@ifx{#1\undefined}
}%
\providecommand \@ifnum [1]{%
 \ifnum #1\expandafter \@firstoftwo
 \else \expandafter \@secondoftwo
 \fi
}%
\providecommand \@ifx [1]{%
 \ifx #1\expandafter \@firstoftwo
 \else \expandafter \@secondoftwo
 \fi
}%
\providecommand \natexlab [1]{#1}%
\providecommand \enquote  [1]{``#1''}%
\providecommand \bibnamefont  [1]{#1}%
\providecommand \bibfnamefont [1]{#1}%
\providecommand \citenamefont [1]{#1}%
\providecommand \href@noop [0]{\@secondoftwo}%
\providecommand \href [0]{\begingroup \@sanitize@url \@href}%
\providecommand \@href[1]{\@@startlink{#1}\@@href}%
\providecommand \@@href[1]{\endgroup#1\@@endlink}%
\providecommand \@sanitize@url [0]{\catcode `\\12\catcode `\$12\catcode
  `\&12\catcode `\#12\catcode `\^12\catcode `\_12\catcode `\%12\relax}%
\providecommand \@@startlink[1]{}%
\providecommand \@@endlink[0]{}%
\providecommand \url  [0]{\begingroup\@sanitize@url \@url }%
\providecommand \@url [1]{\endgroup\@href {#1}{\urlprefix }}%
\providecommand \urlprefix  [0]{URL }%
\providecommand \Eprint [0]{\href }%
\providecommand \doibase [0]{http://dx.doi.org/}%
\providecommand \selectlanguage [0]{\@gobble}%
\providecommand \bibinfo  [0]{\@secondoftwo}%
\providecommand \bibfield  [0]{\@secondoftwo}%
\providecommand \translation [1]{[#1]}%
\providecommand \BibitemOpen [0]{}%
\providecommand \bibitemStop [0]{}%
\providecommand \bibitemNoStop [0]{.\EOS\space}%
\providecommand \EOS [0]{\spacefactor3000\relax}%
\providecommand \BibitemShut  [1]{\csname bibitem#1\endcsname}%
\let\auto@bib@innerbib\@empty
\bibitem [{\citenamefont {Zayats}\ \emph {et~al.}(2005)\citenamefont {Zayats},
  \citenamefont {Smolyaninov},\ and\ \citenamefont
  {Maradudin}}]{Zayats2005131}%
  \BibitemOpen
  \bibfield  {author} {\bibinfo {author} {\bibfnamefont {A.~V.}\ \bibnamefont
  {Zayats}}, \bibinfo {author} {\bibfnamefont {I.~I.}\ \bibnamefont
  {Smolyaninov}}, \ and\ \bibinfo {author} {\bibfnamefont {A.~A.}\ \bibnamefont
  {Maradudin}},\ }\href {\doibase
  http://dx.doi.org/10.1016/j.physrep.2004.11.001} {\bibfield  {journal}
  {\bibinfo  {journal} {Phys Rep}\ }\textbf {\bibinfo {volume} {408}},\
  \bibinfo {pages} {131 } (\bibinfo {year} {2005})}\BibitemShut {NoStop}%
\bibitem [{\citenamefont {Stockman}(2011)}]{Stockman:11}%
  \BibitemOpen
  \bibfield  {author} {\bibinfo {author} {\bibfnamefont {M.~I.}\ \bibnamefont
  {Stockman}},\ }\href {\doibase 10.1364/OE.19.022029} {\bibfield  {journal}
  {\bibinfo  {journal} {Opt. Express}\ }\textbf {\bibinfo {volume} {19}},\
  \bibinfo {pages} {22029} (\bibinfo {year} {2011})}\BibitemShut {NoStop}%
\bibitem [{\citenamefont {Halas}\ \emph {et~al.}(2011)\citenamefont {Halas},
  \citenamefont {Lal}, \citenamefont {Chang}, \citenamefont {Link},\ and\
  \citenamefont {Nordlander}}]{doi:10.1021/cr200061k}%
  \BibitemOpen
  \bibfield  {author} {\bibinfo {author} {\bibfnamefont {N.~J.}\ \bibnamefont
  {Halas}}, \bibinfo {author} {\bibfnamefont {S.}~\bibnamefont {Lal}}, \bibinfo
  {author} {\bibfnamefont {W.-S.}\ \bibnamefont {Chang}}, \bibinfo {author}
  {\bibfnamefont {S.}~\bibnamefont {Link}}, \ and\ \bibinfo {author}
  {\bibfnamefont {P.}~\bibnamefont {Nordlander}},\ }\href {\doibase
  10.1021/cr200061k} {\bibfield  {journal} {\bibinfo  {journal} {Chemical
  Reviews}\ }\textbf {\bibinfo {volume} {111}},\ \bibinfo {pages} {3913}
  (\bibinfo {year} {2011})},\ \bibinfo {note} {pMID: 21542636}\BibitemShut
  {NoStop}%
\bibitem [{\citenamefont {Chikkaraddy}\ \emph {et~al.}(2016)\citenamefont
  {Chikkaraddy}, \citenamefont {de~Nijs}, \citenamefont {Benz}, \citenamefont
  {Barrow}, \citenamefont {Scherman}, \citenamefont {Rosta}, \citenamefont
  {Demetriadou}, \citenamefont {Fox}, \citenamefont {Hess},\ and\ \citenamefont
  {Baumberg}}]{Chikkaraddy:2016aa}%
  \BibitemOpen
  \bibfield  {author} {\bibinfo {author} {\bibfnamefont {R.}~\bibnamefont
  {Chikkaraddy}}, \bibinfo {author} {\bibfnamefont {B.}~\bibnamefont
  {de~Nijs}}, \bibinfo {author} {\bibfnamefont {F.}~\bibnamefont {Benz}},
  \bibinfo {author} {\bibfnamefont {S.~J.}\ \bibnamefont {Barrow}}, \bibinfo
  {author} {\bibfnamefont {O.~A.}\ \bibnamefont {Scherman}}, \bibinfo {author}
  {\bibfnamefont {E.}~\bibnamefont {Rosta}}, \bibinfo {author} {\bibfnamefont
  {A.}~\bibnamefont {Demetriadou}}, \bibinfo {author} {\bibfnamefont
  {P.}~\bibnamefont {Fox}}, \bibinfo {author} {\bibfnamefont {O.}~\bibnamefont
  {Hess}}, \ and\ \bibinfo {author} {\bibfnamefont {J.~J.}\ \bibnamefont
  {Baumberg}},\ }\href {http://dx.doi.org/10.1038/nature17974} {\bibfield
  {journal} {\bibinfo  {journal} {Nature}\ }\textbf {\bibinfo {volume} {535}},\
  \bibinfo {pages} {127} (\bibinfo {year} {2016})}\BibitemShut {NoStop}%
\bibitem [{\citenamefont {Torma}\ and\ \citenamefont {Barnes}(2015)}]{Torma}%
  \BibitemOpen
  \bibfield  {author} {\bibinfo {author} {\bibfnamefont {P.}~\bibnamefont
  {Torma}}\ and\ \bibinfo {author} {\bibfnamefont {W.~L.}\ \bibnamefont
  {Barnes}},\ }\href {http://stacks.iop.org/0034-4885/78/i=1/a=013901}
  {\bibfield  {journal} {\bibinfo  {journal} {Rep. Prog. Phys.}\ }\textbf
  {\bibinfo {volume} {78}},\ \bibinfo {pages} {013901} (\bibinfo {year}
  {2015})}\BibitemShut {NoStop}%
\bibitem [{\citenamefont {Murray}\ and\ \citenamefont
  {Barnes}(2007)}]{ADMA:ADMA200700678}%
  \BibitemOpen
  \bibfield  {author} {\bibinfo {author} {\bibfnamefont {W.~A.}\ \bibnamefont
  {Murray}}\ and\ \bibinfo {author} {\bibfnamefont {W.~L.}\ \bibnamefont
  {Barnes}},\ }\href {\doibase 10.1002/adma.200700678} {\bibfield  {journal}
  {\bibinfo  {journal} {Adv. Mater.}\ }\textbf {\bibinfo {volume} {19}},\
  \bibinfo {pages} {3771} (\bibinfo {year} {2007})}\BibitemShut {NoStop}%
\bibitem [{\citenamefont {Kauranen}\ and\ \citenamefont
  {Zayats}(2012)}]{Kauranen:2012aa}%
  \BibitemOpen
  \bibfield  {author} {\bibinfo {author} {\bibfnamefont {M.}~\bibnamefont
  {Kauranen}}\ and\ \bibinfo {author} {\bibfnamefont {A.~V.}\ \bibnamefont
  {Zayats}},\ }\href {http://dx.doi.org/10.1038/nphoton.2012.244} {\bibfield
  {journal} {\bibinfo  {journal} {Nat Photon}\ }\textbf {\bibinfo {volume}
  {6}},\ \bibinfo {pages} {737} (\bibinfo {year} {2012})}\BibitemShut {NoStop}%
\bibitem [{\citenamefont {Butet}\ \emph {et~al.}(2015)\citenamefont {Butet},
  \citenamefont {Brevet},\ and\ \citenamefont {Martin}}]{Butet:2015aa}%
  \BibitemOpen
  \bibfield  {author} {\bibinfo {author} {\bibfnamefont {J.}~\bibnamefont
  {Butet}}, \bibinfo {author} {\bibfnamefont {P.-F.}\ \bibnamefont {Brevet}}, \
  and\ \bibinfo {author} {\bibfnamefont {O.~J.~F.}\ \bibnamefont {Martin}},\
  }\href {\doibase 10.1021/acsnano.5b04373} {\bibfield  {journal} {\bibinfo
  {journal} {ACS Nano}\ }\textbf {\bibinfo {volume} {9}},\ \bibinfo {pages}
  {10545} (\bibinfo {year} {2015})}\BibitemShut {NoStop}%
\bibitem [{\citenamefont {Mukamel}(1999)}]{MukamelBook}%
  \BibitemOpen
  \bibfield  {author} {\bibinfo {author} {\bibfnamefont {S.}~\bibnamefont
  {Mukamel}},\ }\href@noop {} {\emph {\bibinfo {title} {Principles of Nonlinear
  Optical Spectroscopy}}}\ (\bibinfo  {publisher} {Oxford University Press, New
  York},\ \bibinfo {year} {1999})\BibitemShut {NoStop}%
\bibitem [{\citenamefont {Vasa}\ \emph {et~al.}(2013)\citenamefont {Vasa},
  \citenamefont {Wang}, \citenamefont {Pomraenke}, \citenamefont {Lammers},
  \citenamefont {Maiuri}, \citenamefont {Manzoni}, \citenamefont {Cerullo},\
  and\ \citenamefont {Lienau}}]{Vasa2013}%
  \BibitemOpen
  \bibfield  {author} {\bibinfo {author} {\bibfnamefont {P.}~\bibnamefont
  {Vasa}}, \bibinfo {author} {\bibfnamefont {W.}~\bibnamefont {Wang}}, \bibinfo
  {author} {\bibfnamefont {R.}~\bibnamefont {Pomraenke}}, \bibinfo {author}
  {\bibfnamefont {M.}~\bibnamefont {Lammers}}, \bibinfo {author} {\bibfnamefont
  {M.}~\bibnamefont {Maiuri}}, \bibinfo {author} {\bibfnamefont
  {C.}~\bibnamefont {Manzoni}}, \bibinfo {author} {\bibfnamefont
  {G.}~\bibnamefont {Cerullo}}, \ and\ \bibinfo {author} {\bibfnamefont
  {C.}~\bibnamefont {Lienau}},\ }\href
  {http://dx.doi.org/10.1038/nphoton.2012.340} {\bibfield  {journal} {\bibinfo
  {journal} {Nat Photon}\ }\textbf {\bibinfo {volume} {7}},\ \bibinfo {pages}
  {128} (\bibinfo {year} {2013})}\BibitemShut {NoStop}%
\bibitem [{\citenamefont {Sukharev}\ \emph {et~al.}(2013)\citenamefont
  {Sukharev}, \citenamefont {Seideman}, \citenamefont {Gordon}, \citenamefont
  {Salomon},\ and\ \citenamefont {Prior}}]{Sukharev2013}%
  \BibitemOpen
  \bibfield  {author} {\bibinfo {author} {\bibfnamefont {M.}~\bibnamefont
  {Sukharev}}, \bibinfo {author} {\bibfnamefont {T.}~\bibnamefont {Seideman}},
  \bibinfo {author} {\bibfnamefont {R.~J.}\ \bibnamefont {Gordon}}, \bibinfo
  {author} {\bibfnamefont {A.}~\bibnamefont {Salomon}}, \ and\ \bibinfo
  {author} {\bibfnamefont {Y.}~\bibnamefont {Prior}},\ }\href {\doibase
  10.1021/nn4054528} {\bibfield  {journal} {\bibinfo  {journal} {ACS Nano}\
  }\textbf {\bibinfo {volume} {8}},\ \bibinfo {pages} {807} (\bibinfo {year}
  {2013})}\BibitemShut {NoStop}%
\bibitem [{\citenamefont {Metzger}\ \emph {et~al.}(2016)\citenamefont
  {Metzger}, \citenamefont {Hentschel},\ and\ \citenamefont
  {Giessen}}]{Metzger:2016aa}%
  \BibitemOpen
  \bibfield  {author} {\bibinfo {author} {\bibfnamefont {B.}~\bibnamefont
  {Metzger}}, \bibinfo {author} {\bibfnamefont {M.}~\bibnamefont {Hentschel}},
  \ and\ \bibinfo {author} {\bibfnamefont {H.}~\bibnamefont {Giessen}},\ }\href
  {\doibase 10.1021/acsphotonics.5b00587} {\bibfield  {journal} {\bibinfo
  {journal} {ACS Photonics}\ }\textbf {\bibinfo {volume} {3}},\ \bibinfo
  {pages} {1336} (\bibinfo {year} {2016})}\BibitemShut {NoStop}%
\bibitem [{\citenamefont {Allen}\ and\ \citenamefont {Eberly}(1975)}]{Eberly}%
  \BibitemOpen
  \bibfield  {author} {\bibinfo {author} {\bibfnamefont {L.}~\bibnamefont
  {Allen}}\ and\ \bibinfo {author} {\bibfnamefont {J.~H.}\ \bibnamefont
  {Eberly}},\ }\href@noop {} {\emph {\bibinfo {title} {Optical Resonance and
  Two-Level Atoms}}}\ (\bibinfo  {publisher} {John Wiley \& Sons, Inc.},\
  \bibinfo {year} {1975})\BibitemShut {NoStop}%
\bibitem [{\citenamefont {Abella}\ \emph {et~al.}(1966)\citenamefont {Abella},
  \citenamefont {Kurnit},\ and\ \citenamefont {Hartmann}}]{PhysRev.141.391}%
  \BibitemOpen
  \bibfield  {author} {\bibinfo {author} {\bibfnamefont {I.~D.}\ \bibnamefont
  {Abella}}, \bibinfo {author} {\bibfnamefont {N.~A.}\ \bibnamefont {Kurnit}},
  \ and\ \bibinfo {author} {\bibfnamefont {S.~R.}\ \bibnamefont {Hartmann}},\
  }\href {\doibase 10.1103/PhysRev.141.391} {\bibfield  {journal} {\bibinfo
  {journal} {Phys. Rev.}\ }\textbf {\bibinfo {volume} {141}},\ \bibinfo {pages}
  {391} (\bibinfo {year} {1966})}\BibitemShut {NoStop}%
\bibitem [{\citenamefont {Mandel}\ and\ \citenamefont {Wolf}(1995)}]{Mandel}%
  \BibitemOpen
  \bibfield  {author} {\bibinfo {author} {\bibfnamefont {L.}~\bibnamefont
  {Mandel}}\ and\ \bibinfo {author} {\bibfnamefont {E.}~\bibnamefont {Wolf}},\
  }\href@noop {} {\emph {\bibinfo {title} {Optical Coherence and Quantum
  Optics}}}\ (\bibinfo  {publisher} {Cambridge University Press},\ \bibinfo
  {year} {1995})\BibitemShut {NoStop}%
\bibitem [{\citenamefont {de~Boeij}\ \emph {et~al.}(1998)\citenamefont
  {de~Boeij}, \citenamefont {Pshenichnikov},\ and\ \citenamefont
  {Wiersma}}]{Boeij1998}%
  \BibitemOpen
  \bibfield  {author} {\bibinfo {author} {\bibfnamefont {W.~P.}\ \bibnamefont
  {de~Boeij}}, \bibinfo {author} {\bibfnamefont {M.~S.}\ \bibnamefont
  {Pshenichnikov}}, \ and\ \bibinfo {author} {\bibfnamefont {D.~A.}\
  \bibnamefont {Wiersma}},\ }\href {\doibase 10.1146/annurev.physchem.49.1.99}
  {\bibfield  {journal} {\bibinfo  {journal} {Annu. Rev. Phys. Chem.}\ }\textbf
  {\bibinfo {volume} {49}},\ \bibinfo {pages} {99} (\bibinfo {year}
  {1998})}\BibitemShut {NoStop}%
\bibitem [{\citenamefont {Cho}\ \emph {et~al.}(1992)\citenamefont {Cho},
  \citenamefont {Scherer}, \citenamefont {Fleming},\ and\ \citenamefont
  {Mukamel}}]{Cho1992}%
  \BibitemOpen
  \bibfield  {author} {\bibinfo {author} {\bibfnamefont {M.}~\bibnamefont
  {Cho}}, \bibinfo {author} {\bibfnamefont {N.~F.}\ \bibnamefont {Scherer}},
  \bibinfo {author} {\bibfnamefont {G.~R.}\ \bibnamefont {Fleming}}, \ and\
  \bibinfo {author} {\bibfnamefont {S.}~\bibnamefont {Mukamel}},\ }\href
  {\doibase http://dx.doi.org/10.1063/1.462686} {\bibfield  {journal} {\bibinfo
   {journal} {J. Chem. Phys.}\ }\textbf {\bibinfo {volume} {96}},\ \bibinfo
  {pages} {5618} (\bibinfo {year} {1992})}\BibitemShut {NoStop}%
\bibitem [{\citenamefont {Langer}\ \emph {et~al.}(2014)\citenamefont {Langer},
  \citenamefont {V.}, \citenamefont {A.}, \citenamefont {Salewski},
  \citenamefont {R.}, \citenamefont {Karczewski}, \citenamefont {Wojtowicz},
  \citenamefont {A.},\ and\ \citenamefont {Bayer}}]{Langer2014}%
  \BibitemOpen
  \bibfield  {author} {\bibinfo {author} {\bibfnamefont {L.}~\bibnamefont
  {Langer}}, \bibinfo {author} {\bibfnamefont {S.}~\bibnamefont {V.},
  \bibfnamefont {Poltavtsev}}, \bibinfo {author} {\bibfnamefont
  {Y.}~\bibnamefont {A.}}, \bibinfo {author} {\bibfnamefont {M.}~\bibnamefont
  {Salewski}}, \bibinfo {author} {\bibfnamefont {Y.~D.}\ \bibnamefont {R.}},
  \bibinfo {author} {\bibfnamefont {G.}~\bibnamefont {Karczewski}}, \bibinfo
  {author} {\bibfnamefont {T.}~\bibnamefont {Wojtowicz}}, \bibinfo {author}
  {\bibfnamefont {A.}~\bibnamefont {A.}}, \ and\ \bibinfo {author}
  {\bibfnamefont {M.}~\bibnamefont {Bayer}},\ }\href
  {http://dx.doi.org/10.1038/nphoton.2014.219} {\bibfield  {journal} {\bibinfo
  {journal} {Nat Photon}\ }\textbf {\bibinfo {volume} {8}},\ \bibinfo {pages}
  {851} (\bibinfo {year} {2014})}\BibitemShut {NoStop}%
\bibitem [{\citenamefont {Taflove}\ and\ \citenamefont
  {Hagness}(2005)}]{Taflove}%
  \BibitemOpen
  \bibfield  {author} {\bibinfo {author} {\bibfnamefont {A.}~\bibnamefont
  {Taflove}}\ and\ \bibinfo {author} {\bibfnamefont {S.}~\bibnamefont
  {Hagness}},\ }\href@noop {} {\emph {\bibinfo {title} {Computational
  Electrodynamics: The Finite-Difference Time-Domain Method}}}\ (\bibinfo
  {publisher} {Artech House},\ \bibinfo {year} {2005})\BibitemShut {NoStop}%
\bibitem [{\citenamefont {Gray}\ and\ \citenamefont {Kupka}(2003)}]{Gray2003}%
  \BibitemOpen
  \bibfield  {author} {\bibinfo {author} {\bibfnamefont {S.~K.}\ \bibnamefont
  {Gray}}\ and\ \bibinfo {author} {\bibfnamefont {T.}~\bibnamefont {Kupka}},\
  }\href {\doibase 10.1103/PhysRevB.68.045415} {\bibfield  {journal} {\bibinfo
  {journal} {Phys. Rev. B}\ }\textbf {\bibinfo {volume} {68}},\ \bibinfo
  {pages} {045415} (\bibinfo {year} {2003})}\BibitemShut {NoStop}%
\bibitem [{\citenamefont {Puthumpally-Joseph}\ \emph
  {et~al.}(2015)\citenamefont {Puthumpally-Joseph}, \citenamefont {Atabek},
  \citenamefont {Sukharev},\ and\ \citenamefont
  {Charron}}]{PhysRevA.91.043835}%
  \BibitemOpen
  \bibfield  {author} {\bibinfo {author} {\bibfnamefont {R.}~\bibnamefont
  {Puthumpally-Joseph}}, \bibinfo {author} {\bibfnamefont {O.}~\bibnamefont
  {Atabek}}, \bibinfo {author} {\bibfnamefont {M.}~\bibnamefont {Sukharev}}, \
  and\ \bibinfo {author} {\bibfnamefont {E.}~\bibnamefont {Charron}},\ }\href
  {\doibase 10.1103/PhysRevA.91.043835} {\bibfield  {journal} {\bibinfo
  {journal} {Phys. Rev. A}\ }\textbf {\bibinfo {volume} {91}},\ \bibinfo
  {pages} {043835} (\bibinfo {year} {2015})}\BibitemShut {NoStop}%
\bibitem [{\citenamefont {Puthumpally-Joseph}\ \emph
  {et~al.}(2014)\citenamefont {Puthumpally-Joseph}, \citenamefont {Sukharev},
  \citenamefont {Atabek},\ and\ \citenamefont {Charron}}]{Raiju2014}%
  \BibitemOpen
  \bibfield  {author} {\bibinfo {author} {\bibfnamefont {R.}~\bibnamefont
  {Puthumpally-Joseph}}, \bibinfo {author} {\bibfnamefont {M.}~\bibnamefont
  {Sukharev}}, \bibinfo {author} {\bibfnamefont {O.}~\bibnamefont {Atabek}}, \
  and\ \bibinfo {author} {\bibfnamefont {E.}~\bibnamefont {Charron}},\ }\href
  {\doibase 10.1103/PhysRevLett.113.163603} {\bibfield  {journal} {\bibinfo
  {journal} {Phys. Rev. Lett.}\ }\textbf {\bibinfo {volume} {113}},\ \bibinfo
  {pages} {163603} (\bibinfo {year} {2014})}\BibitemShut {NoStop}%
\bibitem [{\citenamefont {Blake}\ and\ \citenamefont
  {Sukharev}(2015)}]{Blake2015}%
  \BibitemOpen
  \bibfield  {author} {\bibinfo {author} {\bibfnamefont {A.}~\bibnamefont
  {Blake}}\ and\ \bibinfo {author} {\bibfnamefont {M.}~\bibnamefont
  {Sukharev}},\ }\href {\doibase 10.1103/PhysRevB.92.035433} {\bibfield
  {journal} {\bibinfo  {journal} {Phys. Rev. B}\ }\textbf {\bibinfo {volume}
  {92}},\ \bibinfo {pages} {035433} (\bibinfo {year} {2015})}\BibitemShut
  {NoStop}%
\bibitem [{\citenamefont {Weiner}(2009)}]{0034-4885-72-6-064401}%
  \BibitemOpen
  \bibfield  {author} {\bibinfo {author} {\bibfnamefont {J.}~\bibnamefont
  {Weiner}},\ }\href {http://stacks.iop.org/0034-4885/72/i=6/a=064401}
  {\bibfield  {journal} {\bibinfo  {journal} {Rep. Prog. Phys.}\ }\textbf
  {\bibinfo {volume} {72}},\ \bibinfo {pages} {064401} (\bibinfo {year}
  {2009})}\BibitemShut {NoStop}%
\bibitem [{\citenamefont {Weiner}(2011)}]{Weiner:11}%
  \BibitemOpen
  \bibfield  {author} {\bibinfo {author} {\bibfnamefont {J.}~\bibnamefont
  {Weiner}},\ }\href {\doibase 10.1364/OE.19.016139} {\bibfield  {journal}
  {\bibinfo  {journal} {Opt. Express}\ }\textbf {\bibinfo {volume} {19}},\
  \bibinfo {pages} {16139} (\bibinfo {year} {2011})}\BibitemShut {NoStop}%
\bibitem [{\citenamefont {Sukharev}\ and\ \citenamefont
  {Charron}(2016)}]{EricMaxim_molecules}%
  \BibitemOpen
  \bibfield  {author} {\bibinfo {author} {\bibfnamefont {M.}~\bibnamefont
  {Sukharev}}\ and\ \bibinfo {author} {\bibfnamefont {E.}~\bibnamefont
  {Charron}},\ }\href@noop {} {\enquote {\bibinfo {title} {to be published},}\
  } (\bibinfo {year} {2016})\BibitemShut {NoStop}%
\bibitem [{\citenamefont {Brixner}\ and\ \citenamefont
  {Gerber}(2001)}]{Brixner:01}%
  \BibitemOpen
  \bibfield  {author} {\bibinfo {author} {\bibfnamefont {T.}~\bibnamefont
  {Brixner}}\ and\ \bibinfo {author} {\bibfnamefont {G.}~\bibnamefont
  {Gerber}},\ }\href {\doibase 10.1364/OL.26.000557} {\bibfield  {journal}
  {\bibinfo  {journal} {Opt. Lett.}\ }\textbf {\bibinfo {volume} {26}},\
  \bibinfo {pages} {557} (\bibinfo {year} {2001})}\BibitemShut {NoStop}%
\end{thebibliography}
\end{document}